\documentclass[aps,pra,reprint,nofootinbib,showpacs,superscriptaddress]{revtex4-1}

\usepackage[dvipsnames]{xcolor}
\usepackage{units}
\usepackage{amsmath}
\usepackage{float}
\usepackage[colorlinks=true]{hyperref}
\usepackage{amsmath}

\hypersetup{ linkcolor = green,           urlcolor  = blue,            citecolor = blue}
\makeatletter
\usepackage{bm}
\usepackage{amsthm} 
\usepackage{graphicx} 

\let\baraccent=\= 
\renewcommand{\=}[1]{\stackrel{#1}{=}} 

\theoremstyle{definition}

\theoremstyle{remark}

 \def\be{\begin{equation}}
\def\ee{\end{equation}}
\def\bes{\begin{eqnarray}}
\def\ees{\end{eqnarray}}

\begin{document}

\title{Experimental decoy state BB84 quantum key distribution through a turbulent channel}

\author{Eleftherios Moschandreou}
\email{emoschan@vols.utk.edu}
\affiliation{Department of Physics and Astronomy, University of Tennessee, Knoxville,
	TN 37996-1200, USA.}

\author{Brian J. Rollick}
\email{brollick@vols.utk.edu}

\affiliation{Department of Physics and Astronomy, University of Tennessee, Knoxville,
	TN 37996-1200, USA.} 

\author{Bing Qi}
\email{qib1@ornl.gov}
\affiliation{Quantum Information Science Group, Computational Sciences and Engineering Division,
Oak Ridge National Laboratory, Oak Ridge, Tennessee 37831-6418, USA.}

\author{George Siopsis}
\email{siopsis@tennessee.edu}
\affiliation{Department of Physics and Astronomy, University of Tennessee, Knoxville,
	TN 37996-1200, USA.}

\date{\today}
\begin{abstract}
In free-space  Quantum Key Distribution (QKD) in turbulent conditions, scattering and beam wandering cause intensity fluctuations which decrease the detected signal-to-noise ratio. This effect can be mitigated by rejecting received bits when the channel's transmittance is below a threshold. Thus, the overall error rate is reduced and the secure key rate increases despite the deletion of bits. 
In this work, we implement recently proposed selection methods focusing on the Prefixed-Threshold Real-time Selection (P-RTS) where a cutoff can be chosen prior to data collection and independently of the transmittance distribution. We perform finite-size decoy-state BB84 QKD in a laboratory setting where we simulate the atmospheric turbulence using an acousto-optical modulator. We show that P-RTS can yield considerably higher secure key rates for a wide range of the atmospheric channel parameters. In addition, we evaluate the performance of the P-RTS method for a realistically finite sample size. We demonstrate that a near-optimal selection threshold can be predetermined even with imperfect knowledge of the channel transmittance distribution parameters.
\end{abstract}
\maketitle

\section{Introduction}
As quantum communication grows from proof-of-principle laboratory demonstrations towards large-scale commercial
deployment, a lot of attention is focused on the 
optical medium such networks can be realized on.

Today, quantum communication through fiber optical networks is possible at metropolitan scales \cite{Sasaki2011,Tang2016}, but limited in distance due to transmission losses, typically $\sim$ 0.2 dB/km at 1550 nm wavelength \cite{Simon2017}. While classical optical signals can be enhanced by intermediate amplifiers and reach far larger distances, such techniques cannot be employed to amplify quantum signals due to the no-cloning theorem \cite{Wootters1982}. Quantum repeaters \cite{Briegel1998} are a possible solution, but much progress needs to be made before they become available for practical quantum communication. Free-space channels offer an attractive alternative
at intermediate distances for mobile, or remote communicating parties,
or as part of a ground-to-satellite network. So far, experimental demonstrations
in free space include ground-to-airplane \cite{Pugh2017,Nauerth2013}, hot
air balloon \cite{Wang2013}, and drones \cite{Liu2019}, as well as multiple studies on the feasibility of ground-to-satellite quantum communication \cite{Rarity_2002,Meyer-Scott2011,Wang2013,Vallone2015a,Liao2017} and the launch of a QKD dedicated satellite \cite{Liao2017a,Liao2018,Yin2020}.


Signals traveling in free space experience losses due to turbulence, atmospheric absorption and scattering, and consequentially experience consistent degradation of the signal intensity. Caused by fluctuations in the air temperature and pressure,
turbulent eddies of various sizes produce random variations in the atmospheric
refractive index, which cause beam wandering and deformation of the beam front \cite{Vasylyev2016,Vasylyev2018}.

The description of light
propagation in a turbulent medium is a very difficult problem, but
the channel can be described statistically. It is commonly accepted that the transmission coefficient can be approximated by a lognormal probability distribution at moderate turbulence \cite{Diament1970,Milonni_2004,Capraro2012}, and by a gamma-gamma distribution at higher turbulence \cite{Al-Habash2001,Ghassemlooy2012}.  
However, most work to date treats the effect of turbulence on the transmittance as an average loss, without considering the details of the distribution of the transmission coefficient.

Taking the channel statistics into account, various selection methods
that reject or discard recorded bits when the channel transmittance is low have been recently proposed. Evren, \emph{et al.}, \cite{Erven_2012}
developed a signal-to-noise-ratio (SNR) filter where the detected quantum signals are grouped into bins during post-processing. Any bins with
a detection rate below a certain threshold are discarded. To maximize the secure key rate, a searching algorithm was developed to find the optimal bin size and cutoff threshold.

Vallone, \emph{et
al.}, \cite{Vallone2015} employed an auxiliary classical laser beam
to probe the channel statistics, and observed good correlation
between the classical and quantum transmittance data. They developed the 
Adaptive Real-Time Selection (ARTS) method, where the probed channel statistics are used to post-select bits recorded during high transmittance periods, above a certain transmittance threshold. Higher cutoff thresholds improve the SNR at the cost of reducing the number of available signals so the optimal threshold is determined by numerically maximizing the extracted secure key in post-selection. 

Wang, \emph{et al.}, \cite{Wang2018} proposed the Prefixed-Real Time Selection (P-RTS) method and showed that the optimal selection threshold is
insensitive to the channel statistics. Rather, it depends primarily on the
receiver's detection setup characteristics (i.e., the detection efficiency
and background noise) and less strongly on the intensity of the quantum
signals. Thus, the threshold can be predetermined without knowledge of the channel statistics, and the rejection of the recorded bits
can be accomplished in real time without the need to store unnecessary
bits or perform additional post-processing. The P-RTS method was extended to the Measurement-Device-Independent
QKD (MDI QKD)  \cite{Lo2012} protocol in recent studies \cite{Zhu2018,Wang2019}. In particular, ref.\ \cite{Wang2019} highlights the importance of applying a selection method in MDI QKD as the turbulence impacts the protocol's  efficiency not only through the SNR but also through the asymmetry between the channels the communicating parties (Alice and Bob) each use to access the middleman (Charles). 


In this study, the P-RTS method is employed experimentally on the finite-size decoy-state BB84 \cite{Bennett2014,Lim2014} QKD protocol and compared to the optimal key rate found through ARTS. The random transmittance fluctuations caused by the atmospheric turbulence are simulated using an acousto-optical modulator (AOM). We demonstrate that the P-RTS method significantly increases the secure key rate compared to the case of not using post-selection, for a wide range of the channel's parameters. Performing the experiments in a laboratory environment allows the study of different atmospheric conditions in a controllable and reproducible manner and this work extends the array of studies that explore aspects of turbulent Quantum Communication channels with in-lab or simulated methods \cite{Bohmann2017,Rickenstorff2016,Wang2016,Chaiwongkhot2019}.

In Section \ref{sec:2}, we review the features of the P-RTS method \cite{Wang2018}, and discuss how atmospheric effects might alter a free-space communication channel. In Section \ref{sec:ExpSetup}, we describe our experimental setup and procedure. We outline the key generation analysis and present our results in Section \ref{sec:analysis}. Finally, in Section \ref{sec:5} we offer concluding remarks. 

\section{Theory}\label{sec:2}

In this Section, we review the main results of the P-RTS method \cite{Wang2018}, and discuss the atmospheric conditions which might produce our simulated effects.

\subsection{Modeling a Turbulent Atmosphere}

It is accepted that weak to moderate turbulence causes the  transmittance coefficient of light propagating in air, $\eta$, to fluctuate following a lognormal distribution \cite{Osche2002}. The probability density of the transmittance coefficient (PDTC) is given by:
\begin{equation}
\label{eq:lognormal}
    p_{\eta_0,\sigma}(\eta) = \frac{1}{\sqrt{2\pi}\sigma\eta}\exp{\left[-\frac{(\mathrm{ln}(\frac{\eta}{\eta_0})+\frac{\sigma^2}{2})^2}{2\sigma^2}\right]}
\end{equation}
where $\eta_0$ is the average transmittance, and $\sigma^2$ is the log irradiance variance which characterizes the severity of the turbulence. A larger $\sigma^2$ indicates a greater transmittance fluctuation. If the length $L$ of the channel is known and height is constant, $\sigma^2$ for a plane wave can be calculated through the relation: $\sigma^2 = 1.23C_n^2k^{7/6}L^{11/6}$, where $k$ is the wavenumber, and $C_n^2$ is the refractive index structure constant, which could be measured using a scintillometer. Because we are treating the height as a constant, we can assume $C_n^2$ is constant over the channel \cite{Karp2013}. Typical values for $C_n^2$ generally range from $10^{-17}$ to $10^{-12}$ m$^{2/3}$ (going from weak to strong turbulence), with a typical value being $\sim 10^{-15}$ m$^{2/3}$ \cite{Goodman2015}. For example, when $C_n^2 = 5 \times 10^{-15}~\text{m}^{-2/3}$, a value which corresponds to moderate turbulence, we arrive at $\sigma = 0.9$ in a 3 km channel given our 1550 nm wavelength. This choice of $\sigma$ is consistent with prior work (see, e.g., \cite{Vallone2015}). A similar distribution would be produced if one were to choose a longer channel, albeit with less turbulence. Indeed, in \cite{Vallone2015}, $\sigma = .991$ was measured for a 143 km channel.

It should be pointed out that in the case of strong turbulence ($\sigma^2 \gtrsim 1.2$) the lognormal distribution breaks down \cite{Ghassemlooy2012, Osche2002}. Because we argue that P-RTS can predict a cutoff which is largely independent of the PDTC, our findings are also valid in a higher turbulence scenario. 

In addition to turbulence, the beam will be attenuated by the atmosphere. Different software packages such as FASCODE \cite{smith1978fascode} and MODTRAN \cite{berk1987modtran} have been developed to model the atmospheric transmittance as a function of wavelength. In this work, we use MODTRAN to inform our choice of atmospheric loss, because it takes into account a number of different transition lines for many airborne compounds and simulates the effects of a plethora of different aerosols, such as oceanic mist and even volcanic debris \cite{Berk2016}. In our case, a 3 km channel with 13-19 dB of loss can be produced using a Navy Maritime aerosol model where visibility ranges from about 1.8 to 2.5 km, a range which corresponds to light fog or hazy conditions. For comparison, 30 dB of loss was obtained for the much longer channel (143 km) between the Tenerife and La Palma islands in \cite{Capraro2012}.

\subsection{Key Generation in a Turbulent Channel}
Our experimental setup implements a process in which two users, Alice and Bob, are generating a shared secure key to use for their secret communication. Alice is sending phase randomized weak coherent (laser) pulses where her bits are encoded as the polarization state. Bob receives and detects the pulses using single photon avalanche detectors (SPAD). Note that in both the theoretical calculation and the experimental demonstration, while the average channel loss is assumed to be a constant, the channel loss itself fluctuates according to Eq.\ \eqref{eq:lognormal}. This channel model has been widely adopted in free-space QKD.

\subsubsection{Asymptotic case}
Following the discussion of \cite{Wang2018}, to describe the dependence of the secure key generation rate $R$ on the transmittance $\eta$ of the atmospheric channel, we fix all Alice's decoy state parameters as well as all Bob's detection parameters (i.e., his detectors' efficiencies, background noise and optical misalignment). Details on the optimization process are given in Appendix \ref{optimization}. Then the key rate can be written as a single function of the transmittance, $R\left(\eta\right)$. 

The maximum key rate $R_{\text{max}}$ that can be extracted using the channel's statistics is given by the convolution of the PDTC, $p_{\eta_0,\sigma} (\eta)$ in Eq.\ \ref{eq:lognormal} with the rate $R\left(\eta\right)$,
\begin{equation}
R_{\text{max}}=\int_{0}^{1}R\left(\eta\right)p_{\eta_0,\sigma}\left(\eta\right)\text{d}\eta\label{eq:Ratewise}
\end{equation}
While evaluating this integral is challenging in practical
applications, we can set a transmittance threshold $\eta_{TH}$ below
which recorded bits are discarded and keep only a fraction $\int_{\eta_{TH}}^{1}p_{\eta_0,\sigma}\left(\eta\right)\text{d}\eta$
of the sent signals. We treat the remaining recordings as having
passed through a static channel of average transmittance $\left\langle \eta\right\rangle$, computed only from the transmittances above the threshold:
\begin{equation}
\left\langle \eta\right\rangle =\dfrac{\int_{\eta_{TH}}^{1}\eta p_{\eta_0,\sigma}\left(\eta\right)\text{d}\eta}{\int_{\eta_{TH}}^{1}p_{\eta_0,\sigma}\left(\eta\right)\text{d}\eta}\label{eq:truncated}
\end{equation}
Then the postselected bits produce a key rate \cite{Wang2018}: 
\begin{equation}
R\left(\eta_{TH}\right)=R\left(\left\langle \eta\right\rangle \right)\times\int_{\eta_{TH}}^{1}p_{\eta_0,\sigma}\left(\eta\right)\text{d}\eta\label{eq:Rsimplified}
\end{equation}
Eq.\ \eqref{eq:Rsimplified} presents an optimization problem: higher cutoffs $\eta_T$ improve the SNR for the postselected bits and, hence, the rate $R\left(\left\langle \eta\right\rangle \right)$ at the cost of reducing the available signals $\int_{\eta_{TH}}^{1}p_{\eta_0,\sigma}\left(\eta\right)\text{d}\eta$.
The authors of ref.\ \cite{Wang2018} showed that an optimal threshold $\eta_T$ can be predetermined and the resulting key generation rate \eqref{eq:Rsimplified}
can closely approach the ideal rate of Eq.\ \eqref{eq:Ratewise}
by making two key observations. Firstly, there exists a critical transmittance
$\eta_{CR}$ such that $R(\eta)=0$, for $\eta<\eta_{CR}$. Thus, we have
\begin{equation}
R_{\text{max}}=\int_{0}^{1}R\left(\eta\right)p_{\eta_0,\sigma}\left(\eta\right)\text{d}\eta=\int_{\eta_{CR}}^{1}R\left(\eta\right)p_{\eta_0,\sigma}\left(\eta\right)\text{d}\eta\label{eq:CriticalRatewise}
\end{equation}
Secondly, the rate $R\left(\eta\right)$, although convex in general,
approaches linearity very well. Approximating the rate $R\left(\eta\right)$ as linear, $R\left(\eta\right)\approx\alpha\cdot\eta+\beta$,
we have, 
\begin{align}
R_{\text{max}}=&\int_{\eta_{CR}}^{1}R\left(\eta\right)p_{\eta_{0},\sigma}\left(\eta\right)\text{d}\eta \nonumber 
\\
 \approx& \int_{\eta_{CR}}^{1}\alpha\!\cdot\!\eta p_{\eta_{0},\sigma}\left(\eta\right)\text{d}\eta+\int_{\eta_{CR}}^{1}\beta\!\cdot\!p_{\eta_{0},\sigma}\left(\eta\right)\text{d}\eta\nonumber
 \\
 =& R\left(\left\langle \eta\right\rangle \right)\times\int_{\eta_{CR}}^{1}p_{\eta_{0},\sigma}\left(\eta\right)\text{d}\eta\label{eq:final_approx_rate}
\end{align}
This implies that by setting our threshold to the critical value, $\eta_{TH}=\eta_{CR}$, in Eq.\ \eqref{eq:Rsimplified}, we achieve a very good approximation of $R_{\text{max}}$. Importantly, the optimal transmittance cutoff does not depend on the channel's transmittance parameters, $\{\eta_0,\sigma\}$.

\subsubsection{Finite-size effects}

Taking the finite-size effects into consideration, the extracted secure key rate $R_{\text{finite-size}}$ depends also on the number of pulses $N$ sent by Alice. Discarding low transmittance events reduces the available postselected pulses to $N_{\text{post}}=N\times\int_{\eta_{TH}}^{1}p_{\eta_0,\sigma}\left(\eta\right)\text{d}\eta$, so the distilled secure key rate is modified to \cite{Wang2018}:

\begin{equation}
R=R_{\text{finite-size}}\left(\left\langle \eta\right\rangle ,N_{\text{post}}\right)\!\!\times\!\!\!\int_{\eta_{TH}}^{1}p_{\eta_0,\sigma}\left(\eta\right)\text{d}\eta\label{eq:Rfinite}
\end{equation}
The rate $R_{\text{finite-size}}$ is calculated as
\begin{equation}
R_{\text{finite-size}}=\frac{\ell}{N}~,\label{eq:RFinite2}
\end{equation}
where $\ell$ the number of distilled secure bits. The latter is found from
\begin{equation}
\ell = s_{X,0}+s_{X,1}-s_{PA}\left(\phi_X\right)-s_{EC}\left(e_{obs}\right)~,\label{eq:ell}
\end{equation}
where $s_{X,0}$ and $s_{X,1}$ are the contributions from zero and single photon pulses, respectively, and $s_{EC}$  and $s_{PA}$ are the bits consumed to perform error correction and privacy amplification. The contributions $s_{X,0}$ and $s_{X,1}$, as well as the phase error $\phi_X$, are estimated using the two-decoy state method \cite{Ma2005} adapted to include finite-size effects, according to Lim, \emph{et al.}~\cite{Lim2014}. The observed error $e_{obs}$ is measured directly. Details of the secure key rate calculation are presented in Section \ref{sec:analysis}.

\begin{figure}[h]
\begin{center}\includegraphics[scale=1]{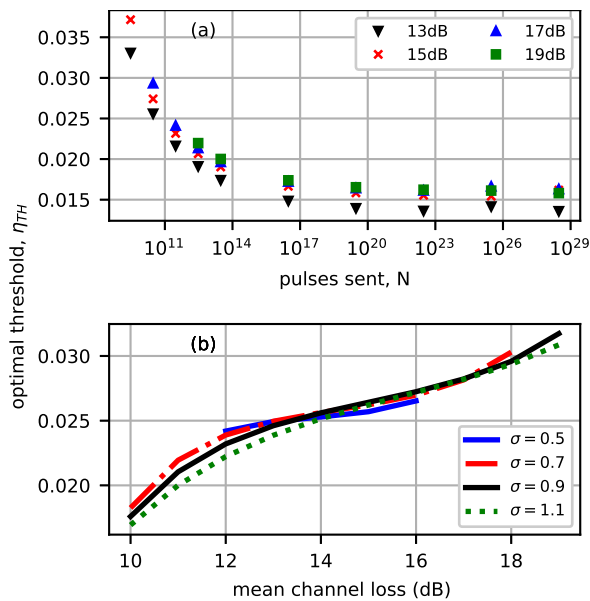}\end{center}
\caption{Simulations. (a) The optimal transmittance threshold, for different number of sent pulses $N$ and for different mean channel losses. Here $\sigma=0.9$ for all simulation points. (b) The optimum threshold in terms of the mean channel loss for different $\sigma$ values. Here $N=3\times 10^{10}$ for all simulation points.
\label{fig:simulations_finite}}
\end{figure}

The dependence of the rate $R_\text{finite-size}$ given by Eq. \eqref{eq:RFinite2} on the number of sent pulses, $N$, raises the question whether the main conclusion of the PRTS method, that the optimum transmittance threshold can be pre-determined independently of the channel statistics, still holds for the case of finite number of sent pulses. Although the form of the distilled bits, $\ell$, (Eqs.\ \eqref{eq:ell} above and \eqref{eq:distilled_ell} in Section \ref{sec:analysis}) does not allow us to easily examine it analytically, we were able to draw conclusions from numerical simulations.

The simulation results are presented in Fig.\ \ref{fig:simulations_finite} for the  parameters presented in Tables \ref{tab:Backgrounds} and \ref{tab:Parameters}. The examined channel loss range (10-20 dB) is the range of most significance for the selection method given our detection parameters. For losses below this range the selection method offers no significant improvement while at greater losses the extracted key rate is still insignificant or zero. The examined $\sigma$ range ($0.5-1.1$) corresponds to typical values found in the literature \cite{Milonni_2004,Capraro2012,Vallone2015}.

Considering that for a realistic application of communication time of a few minutes at frequency 1 GHz, we can send $\sim 10^{11}-10^{12}$ pulses, we observe that the optimum threshold at low number of sent pulses, $N$, may differ from its asymptotic value. We also observe a similar variation on the optimum threshold for different values of the channel's parameters $\eta_0$ and  $\sigma$. Moreover this variation does not affect the secure key generation significantly. Given these observations, we conclude that even with an imperfect knowledge of the channel statistics, we can predetermine a transmittance cutoff which produces a near-optimum key generation rate. We explore this conclusion experimentally in Section \ref{sec:analysis}.

\begin{figure*}[!ht]
\begin{center}
	\includegraphics[scale=0.95]{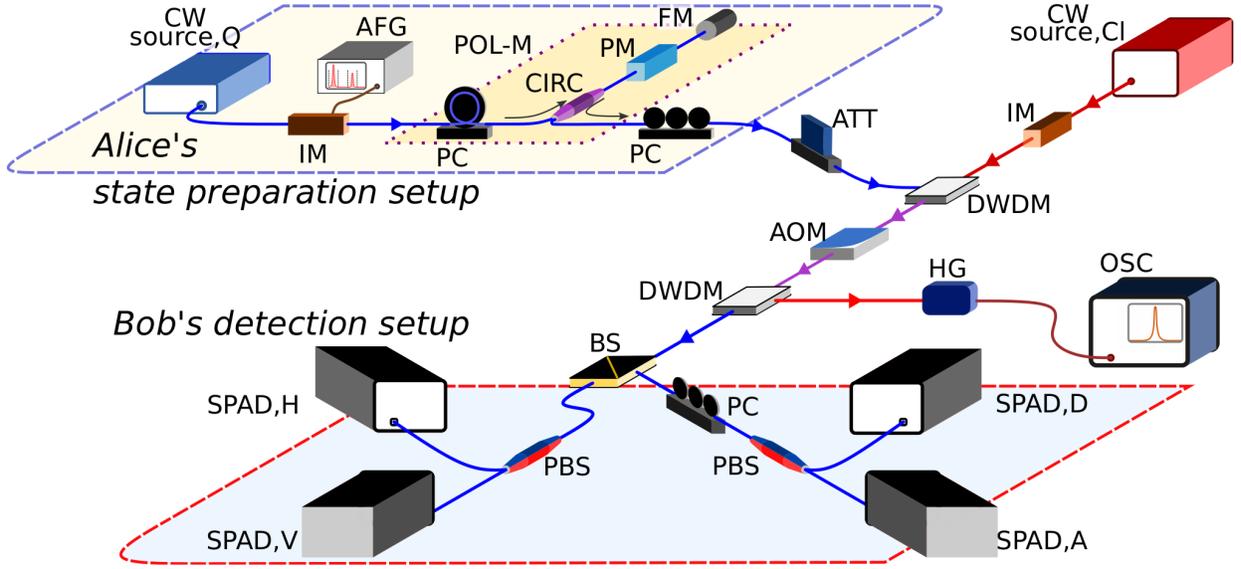}
	\caption{ A continuous wave source (CW source, Q) is used to encode the quantum states on. Pulses are carved with  an intensity modulator (IM) driven by an arbitrary function generator (AFG). The driving voltage sequence contains three voltage scales, implementing the three (signal-weak-vacuum) decoy state method. Bits are encoded as polarization states with a polarization modulation (POL-M) setup consisting of a polarization controller (PC) a circulator (CIRC) a phase modulator (PM)  and a Faraday mirror (FM). A polarization controller is used to align to the rectilinear basis. The beam is attenuated (ATT) to the desired photon number.  An additional laser source (CW source,Cl) with an intensity modulator produces classical pulses that probe the channel statistics. The classical and quantum pulses are multiplexed at a dense wavelength division multiplexer (DWDM). Turbulence is simulated at an acousto-optic modulator (AOM). At Bob's side, a DWDM separates the classical and quantum signals. The high intensity classical beam is read by a high gain (HG) detector and sampled by an oscilloscope (OSC). The quantum beam is sent through a beam splitter (BS) to randomly select the measurement basis. Each polarization, Horizontal, Vertical, Diagonal , Anti-diagonal (H,V,D,A), measurement is realized by a polarization beam splitter (PBS) and a single photon avalanche detector (SPAD). An additional polarization controller is used to align to the Diagonal basis.}
	\label{apparatus}
	\end{center}
\end{figure*}

\section{Experimental Setup}\label{sec:ExpSetup}



The experimental setup is shown in Fig.\ \ref{apparatus}. A continuous-wave (CW) laser source (Wavelength References)
at 1550.5 nm (ITU channel 33.5) is directed to a $\text{LiNbO}_{3}$ (EOSPACE)
intensity modulator (IM) to carve out pulses of full width half maximum  (FWHM) $\sim$ 2 ns at a 25-MHz
repetition rate. The intensity modulator is driven by an arbitrary function generator (AFG) (Tektronix) with a sequence of three different voltage scales, implementing the three-decoy (signal, weak, vacuum) state method. The DC bias voltage of the IM is
automatically adjusted by a Null Point Controller (PlugTech) to achieve
the optimal extinction ratio (typically $\sim$ 30 dB). For each experimental session, Alice prepares and sends $N=3\times10^{10}$ pulses. To implement polarization encoding BB84, we developed a fiber based high-speed polarization modulator, following the design described in \cite{Tang2014} which was proposed in \cite{Lucio-Martinez2009}.

The pulses are attenuated by a combination of digital and analog variable attenuators to single-photon levels. The pulses carrying the quantum states are multiplexed on a dense wavelength division multiplexer (DWDM) (Lightel) with 1554-nm (ITU channel 29) classical laser pulses at 4-kHz repetition rate and $\sim$3 ns  FWHM. The classical pulses are used to probe the channel's transmittance statistics. Both sets of pulses are directed to an AOM (Brimrose) which is used to generate the random transmittance fluctuations expected from our turbulent channel. Another DWDM is employed at the receiver to separate the classical probe light and the quantum signals. The classical laser is detected by a high-gain detector (Thorlabs), and an oscilloscope (Tektronix) is used to sample and store the outputs of the detector. A 50:50 beam splitter (BS) is used to passively select Bob's detection basis, rectilinear or diagonal. Measurement in each basis is realized by a polarizing beam splitter (PBS) and a pair of InGaAs single photon avalanche detectors (SPAD) (IDQ) gated at 25 MHz with $\sim$5-ns gate width.

The detector dead-time is set to 9 $\mu$s to reduce the afterpulse probability.
Since the afterpulse probability depends on the light intensity received
by the detectors, we observe a linear dependence of the background
probability $P_{bg}$ in terms of the channel transmittance $\eta$ of the
form $P_{bg}(\eta)=Y_{0}+b\cdot\eta$. The parameters $Y_{0}$ and
$b$ are extracted experimentally with linear fits from test measurements
and are displayed in Table \ref{tab:Backgrounds} using input light with the same average photon number as that used in the experiments. 

\begin{table}
\begin{tabular}{lcc}
\hline \hline 
 & $Y_{0}$  & $b$\tabularnewline
\hline 
detector H & $\left(7.6\pm0.6\right)\cdot10^{-6}$ & $\left(2.6\pm0.4\right)\cdot10^{-4}$\tabularnewline
detector V & $\left(3.1\pm0.2\right)\cdot10^{-5}$ & $\left(1.8\pm0.4\right)\cdot10^{-4}$\tabularnewline
detector D & $\left(6.7\pm0.3\right)\cdot10^{-5}$ & $\left(2.7\pm0.4\right)\cdot10^{-4}$\tabularnewline
detector A & $\left(6.7\pm0.3\right)\cdot10^{-5}$ & $\left(1.8\pm0.4\right)\cdot10^{-4}$\tabularnewline
\hline \hline 
\end{tabular}
\caption{Background noise parameters for each detector. The input states have the same average photon number as the optimized states given in Table \ref{tab:OptimizedStates}.  The background click probability is given by $P_{bg}(\eta)=Y_0+b\cdot\eta$.\label{tab:Backgrounds}}
\end{table}

The optical misalignment
is approximately $3\times10^{-3}$. Each SPAD is
set to $10\%$ quantum efficiency ($\eta_d$). The experimental parameters
are summarized in Table \ref{tab:Parameters}. Bob's optical efficiency ($\eta_{BOB}$)
refers to losses due to optical components  (i.e. BS, PBS and the fiber links).
The output of each SPAD is recorded by a Time Interval Analyzer (TIA) (IDQ) and
a custom-made program sifts them to collect the sets $nX_{k},mX_{k},nZ_{k},mZ_{k}$,
for $k\in\left\{ \mu_1,\mu_2,\mu_3\right\} $, that are needed
for the secure key distillation parameters according to the model
of \cite{Lim2014}. Here, $nB_{k}$ are the detections
where both Alice and Bob use the same basis $B\in\left\{ X,Z\right\} $
while the decoy intensity $k$ is used, and $mB_{k}$ are the detections
in error for the basis $B$ and decoy intensity $k$. 

\begin{table}[h]
\begin{tabular}{cc}
\hline \hline 
Bob's optical efficiency & $0.42\pm0.02$\tabularnewline
Optical misalignment & $0.003\pm0.002$\tabularnewline
Quantum efficiency (all detectors) & $0.1\pm0.05$\tabularnewline
Dead-time & $9 \mu s$\tabularnewline
$f_{EC}$ & $1.16$\tabularnewline
$N$ & $3\cdot10^{10}$\tabularnewline
\hline \hline 
\end{tabular}
\caption{Experiment parameters.\label{tab:Parameters}}
\end{table}

Given the experimental parameters in Tables \ref{tab:Backgrounds} and \ref{tab:Parameters}, 
we numerically optimize the key generation to find the optimal parameters $\{q_{X},P_{\mu_1},P_{\mu_2},\mu_1,\mu_2\}$. Here,
$q_{X}$ is the probability of using the rectilinear basis, $P_{\mu_1}$
and $P_{\mu_2}$ are the proportions of the signals and weak decoys, and
${\mu_1,\mu_2}$ are the signal and weak decoy intensities for the
desired turbulence parameter set, $\{\eta_{0},\sigma\}$. The vacuum decoy parameters are fixed as $P_{\mu_3}=1-P_{\mu_1}-P_{\mu_2}$, and $\mu_3=0.002$. The optimized states are presented in Table \ref{tab:OptimizedStates} and the details of the optimization routine are presented in Appendix \ref{optimization}.

\begin{table}[h]
\begin{center}%
\begin{tabular}{cccccc}
\hline \hline  
 Turbulence & $q_{x}$ & $P_{\mu_1}$ & $P_{\mu_2}$ & $\mu_1$ & $\mu_2$\tabularnewline
\hline 
$\{\eta_{0}=10^{-1.1},\sigma=0.9\}$ & $0.904$ & $0.660$ & $0.215$ & $0.56$ &
$0.225$\tabularnewline
$\{\eta_{0}=10^{-1.3},\sigma=0.9\}$ & $0.879$ & $0.617$ & $0.244$ & $0.56$ & $0.23$\tabularnewline
$\{\eta_{0}=10^{-1.5},\sigma=0.9\}$ & $0.844$ & $0.552$ & $0.287$ & $0.56$ & $0.23$\tabularnewline
$\{\eta_{0}=10^{-1.7},\sigma=0.9\}$ & $0.789$ & $0.460$ & $0.352$ & $0.54$ & $0.24$\tabularnewline
$\{\eta_{0}=10^{-1.9},\sigma=0.9\}$ & $0.683$ & $0.319$ & $0.439$ & $0.54$ & $0.245$\tabularnewline
\hline \hline 
\end{tabular}\end{center}\caption{Alice's optimized quantum states, for channel parameters $\sigma=0.9$ and mean channel loss {11,13,15,17,19} dB. \label{tab:OptimizedStates}}
\end{table}
\begin{figure*}[!ht]
\begin{center}
\includegraphics[scale=0.95]{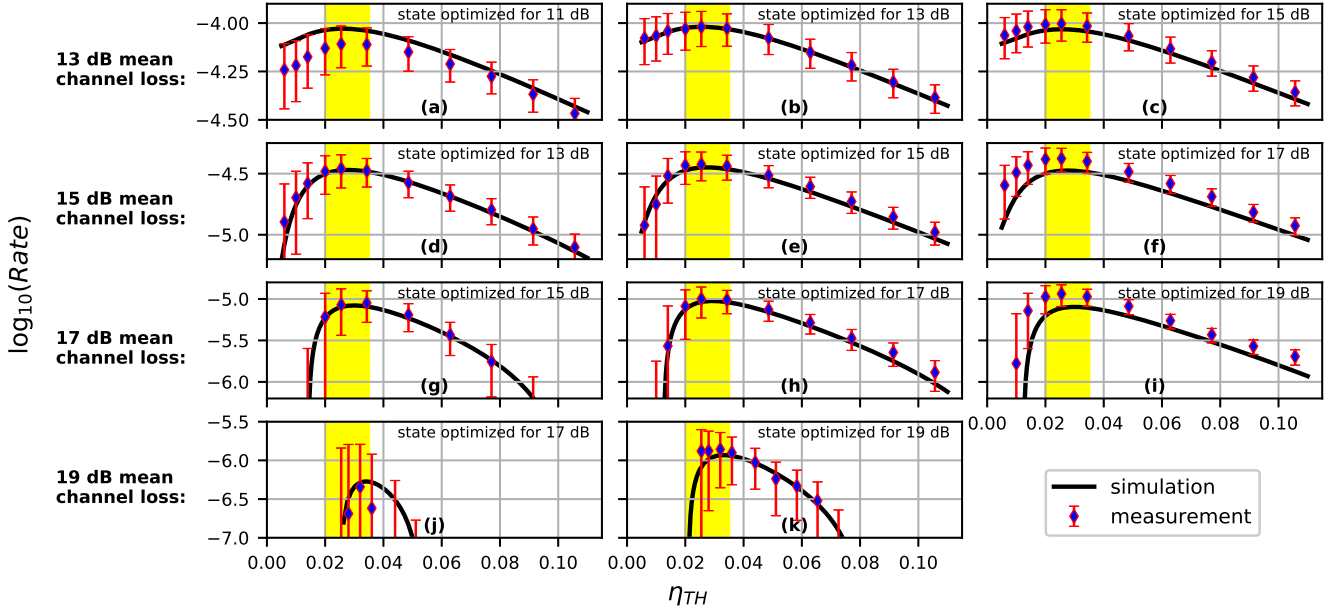}
\caption{ARTS type measurements: the logarithm of the secure key rate for increasing applied transmittance cutoff. First row (a)-(c): Distilled secure key rates at 13 dB mean channel loss for increasing applied transmittance cutoffs. Alice optimizes her state assuming (a) 11 dB, (b) 13 dB, (c) 15 dB mean channel loss. Second row (d)-(f): Similar for 15 dB channel loss and (d) 13 dB, (e) 15 dB, (f) 17 dB loss assumed by Alice. Third row (g)-(i): Similar for 17 dB channel loss and (d) 15 dB, (e) 17 dB, (f) 19 dB loss assumed by Alice. Fourth row (j),(k): 19 dB channel loss and (j) 17 dB, (k) 19 dB loss assumed by Alice, no key is generated if Alice assumes 21 dB loss. The errorbars represent a $\pm 0.005$ uncertainty in the signal and weak decoy average photon number. The yellow shaded areas represent the range of variation on the optimal transmittance threshold, presented in Fig. \ref{fig:simulations_finite}. }
\label{fig:Results13_15_17_19}
\end{center}
\end{figure*}
\section{Analysis and Results\label{sec:analysis}}

Having collected all the sets $nX_{k},mX_{k},nZ_{k},mZ_{k}$ defined in the previous section
for $k\in\left\{ \mu_1,\mu_2,\mu_3\right\} $, we distill according to \cite{Lim2014} a secure key of length $\ell$,
\begin{eqnarray}
\ell & = & \left\lfloor \vphantom{\frac{2}{\varepsilon_{cor}}} s_{X,0}+s_{X,1}\left(1-h(\phi_{X})\right)\right.\label{eq:distilled_ell}\nonumber\\
 &  & \left.-n_{X} \! \cdot \! f_{EC} \! \cdot \!  h(e_{obs})-6\log_{2}\frac{21}{\varepsilon_{sec}}-\log_{2}\frac{2}{\varepsilon_{cor}}\right\rfloor \ \ \ \ \
\end{eqnarray}
where $s_{X,0}$ and $s_{X,1}$ are the lower bounds on the number of bits generated by zero and single photon pulses (which are immune to photon number splitting attacks) while both Alice and Bob use the rectilinear basis, $\phi_{X}$ is the upper bound on the phase error, $h(\cdot)$ is the binary entropy function,
\begin{equation}
h(x)=-x\log_{2}x-\left(1-x\right)\log_{2}\left(1-x\right)    
\end{equation}
and  $e_{obs}=\frac{m_{X}}{n_{X}}$ is the quantum bit error rate, with $m_{X}=mX_{\mu_1}+mX_{\mu_2}+mX_{\mu_3}$, and $n_{X}=nX_{\mu_1}+nX_{\mu_2}+nX_{\mu_3}$.
The term $-n_{X} \! \cdot \! f_{EC} \! \cdot \!  h(e_{obs})$ describes
the bits consumed by the classical error correction algorithm \cite{Brassard1994} 
with efficiency $f_{EC}=1.16$,  and $\varepsilon_{cor}=10^{-15}$ is the correctness parameter. 
The term $-s_{X,1} \! \cdot \! h(\phi_{X})$ describes the bits consumed during the privacy amplification stage to achieve secrecy according to the secrecy parameter $\varepsilon_{sec}=10^{-10}$.

We explore the premise of section \ref{sec:2}, whereby one can predetermine a near optimal transmittance cutoff while considering finite-size effects even with an imperfect knowledge of the channel statistics. For each examined channel loss (11-19 dB), Alice prepares her state parameters while (i) having perfect knowledge of the channel, (ii) underestimating the mean loss by 2 dB, (iii) overestimating the mean loss by 2 dB. For example at 17 dB mean channel loss, Alice assumes (i) 17 dB, (ii) 15 dB, (iii) 19 dB mean channel loss and prepares her state, Table \ref{tab:OptimizedStates}, accordingly. A 2-dB uncertainty window for the mean channel loss can be comfortably achieved by classical means during the initial calibration stage. In any case, such knowledge of the channel parameters is required and should be pursued for the construction of Alice's state (Table \ref{tab:OptimizedStates}) as the state parameters (especially the proportions $\{q_X,P_{\mu_1},P_{\mu_2}\}$) are sensitive to the mean channel loss. For this reason, we did not consider the case of larger uncertainty on the channel loss.

We present our measurement results in Fig.\ \ref{fig:Results13_15_17_19}, for each channel loss and optimized state. The measurement data points correspond to ARTS \cite{Vallone2015}-type post-selection where we scan successive transmittance cutoffs and extract the corresponding secure key rate. The yellow shaded area corresponds to the variance on the optimal cutoff observed in Fig.\ \ref{fig:simulations_finite}. The errorbars represent an uncertainty $\pm 0.005$ in setting during the experiment the desired signal photon number $\mu_1$ and weak decoy photon number $\mu_2$ given in Table \ref{tab:OptimizedStates}. In practical applications though, intensity uncertainties should be treated more formally with  methods such as those discussed in ref.\ \cite{Mizutani2015}.

We observe that for a wide range of mean channel losses (13-17 dB) and within the range of uncertainty of the optimal threshold (yellow shaded range), the extracted secure key rate does not vary significantly from its optimal value. However this conclusion does not hold well at higher losses, where more precise knowledge on the channel parameters is required, in order to both prepare Alice's state parameters and apply the selection threshold.

\begin{figure}[ht!]
\begin{center}\includegraphics[scale=1.0]{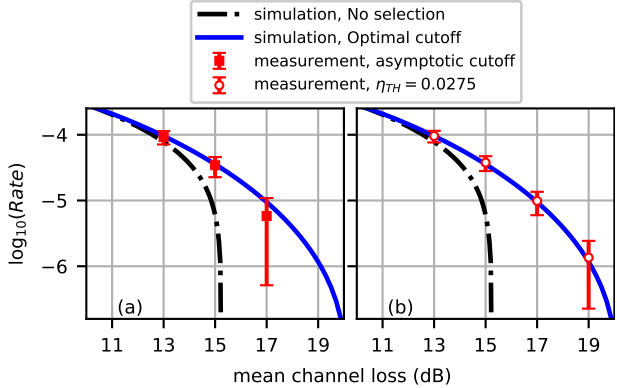}
\end{center}
\caption{P-RTS type measurements. (a) Cutoff set as $\eta_{TH}=0.016$ the asymptotic optimal cutoff. This cutoff does not generate a key at 19 dB mean channel loss (b) Cutoff set as $\eta_{TH}=0.0275$, an average cutoff observed in Fig. \ref{fig:simulations_finite} .\label{fig:PRTSvsNOPRTS}}
\end{figure}

In Fig.\ \ref{fig:PRTSvsNOPRTS} we present an evaluation of our P-RTS type measurements where a fixed and predetermined cutoff transmittance is set. In Fig.\ \ref{fig:simulations_finite}(a) we have observed that the optimal threshold approaches the value $\eta_{TH}=0.016$ as the number of sent pulses N becomes large. We acquire a similar value from the root of the equation $R_{GLLP}(\eta)=0$ where $R_{GLLP}(\eta)$  is the GLLP \cite{Gottesman2002} asymptotic secure key rate as a function of the channel transmittance. This value is the asymptotic cutoff applied in Fig.\ \ref{fig:PRTSvsNOPRTS}(a) and it can be predetermined without any knowledge of the channel statistics \cite{Wang2018}. In Fig.\ \ref{fig:simulations_finite}(b) we have observed a limited variance in the optimal threshold around the value $\eta_{TH}=0.0275$ throughout the range of channel parameters where the selection method offers significant improvement on the extracted key rate. We choose this value to represent a threshold acquired through partial knowledge of the channel parameters.

Fig.\ \ref{fig:PRTSvsNOPRTS} shows that for a wide range of the examined mean channel loss, the asymptotic threshold only slightly under-performs the threshold acquired through partial knowledge of the channel. However at higher losses, a PRTS-type (channel-independent) threshold fails to produce a secure key rate and some partial knowledge on the channel parameters is required. In any case, choosing the asymptotic threshold still allows ARTS-type scanning during post-selection to fully maximize the generated secure key rate.

\section{Concluding remarks}\label{sec:5}
\begin{figure}[ht!]
\begin{center}\includegraphics[scale=0.99]{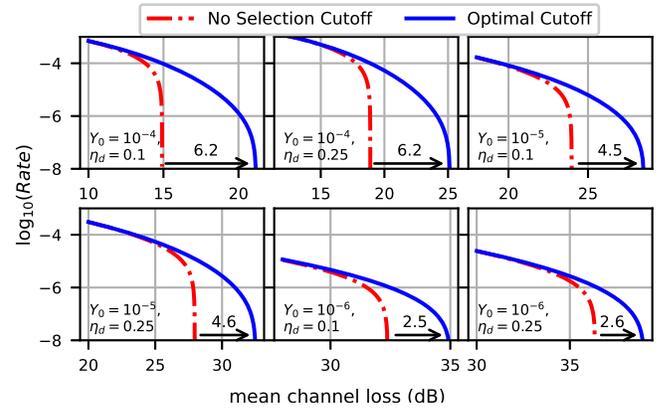}
\end{center}
\caption{Simulations: The improvement that the selection method offers for typical SPAD background ($Y_0$) and quantum efficiency ($\eta_d$) values. As a comparison criterion, we examine the mean channel loss at which a key rate of $10^{-8}$ can be achieved. Sample size is $N=3\cdot10^{10}$, and the optimal cutoff is applied. $\sigma=0.9$.\label{fig:Improvement}}
\end{figure}

We conducted an experimental demonstration of Decoy State BB84 QKD over a simulated turbulent channel taking finite-size effects into account. We showed that the main conclusion of the Prefixed Real-Time Selection (P-RTS) scheme proposed in \cite{Wang2018}, that the transmittance threshold can be predetermined independently of the channel statistics, holds well in the regime of realistically finite events, further supporting the applicability of the method. The secure key rate can be significantly improved in turbulent atmospheric conditions, especially at high loss. The selection method can be easily implemented without any significant technological upgrades, while saving computational resources. We observe that it is especially beneficial for lower quality detection setups, with higher detection noise, as the turbulence impacts their SNR more severely. We offer supporting simulations in Fig.\ \ref{fig:Improvement}. For example, by applying a transmittance cutoff at background noise $Y_0=10^{-4}$,  we can extend the mean channel loss so that a $10^{-8}$ key rate is generated by 6.2 dB. For $Y_0=10^{-6}$, this extension is for 2.5 dB.

Depending on the knowledge of the turbulence statistics, the two selection methods could be used in combination. One could select a conservative transmittance threshold to perform PRTS-type real time data rejection and then perform an ARTS-type scan during postselection to further maximize the extracted secure key rate.

It should be pointed out that one important assumption behind the security proof adopted in this work is that the global phase of Alice's quantum state signal is random \cite{Lo2006}. This could be achieved by using a PM at Alice's station to actively randomize the phase of each quantum signal, as demonstrated in \cite{Zhao2007}. For simplicity, we did not implement phase randomization. Nevertheless, since the coherence time of Alice's laser is much smaller than the data collection time, the detection statistics observed in our experiment match the case where phase randomization is applied.

A similar technique could be used to enhance free-space MDI QKD, as well as other free space protocols where the secure key rate can be approximated as a straight line at the lower boundary. Being able to overcome the challenges of atmospheric turbulence is a crucial step in building a future global quantum network. 

\acknowledgments
We thank Hoi-Kwong Lo and Wenyuan Wang for useful comments, and Raphael Pooser for help with the initial setup. This work was supported by the U.S.\ Office of Naval Research under award number N00014-15-1-2646. Bing Qi acknowledges support from the U.S.\ Department of Energy Office of Cybersecurity Energy Security and Emergency Response (CESER) through the Cybersecurity for Energy Delivery Systems (CEDS) program. 
%

\appendix
\section{Optimizing the Secure Key Rate \label{optimization}}

Decoy state QKD introduces additional degrees of freedom for the pulses sent. Optimization of these parameters can have a profound effect on the secure key rate. In this appendix, we explain how the secure key rate is calculated and describe the optimization process. 
We assume that Alice has full knowledge of Bob's detection setup parameters,
as summarized in Table \ref{tab:Parameters}. She knows that Bob
will apply a selection threshold and she also has some knowledge on the channel
parameters $\left\{ \eta_{0},\sigma\right\}$ according to the discussion in Section \ref{sec:analysis}. For the rest of the
section, we follow the notation of Lim, \emph{et al.}~\cite{Lim2014}
, where $X$ denotes the rectilinear (computational) basis and $Z$
the diagonal (Hadamard) basis. Alice performs a numerical optimization
over the free parameters of her state, $\left\{ q_{X},P_{\mu_1},P_{\mu_2},\mu_1,\mu_2\right\} $,
where $q_{X}$ is the fraction of bits encoded in the X-basis
, $P_{\mu_1}$ and $P_{\mu_2}$ are the fractions of signal
state and weak decoy state bits, respectively, $\mu_1$ and $\mu_2$
are the photon numbers per pulse for the signal and weak decoy states,
respectively. For the vacuum decoy state we have fixed $\mu_3=0.002$,
and $P_{\mu_3}=1-P_{\mu_1}-P_{\mu_2}$.

The detection probability for the decoy $k\in\left\{ \text{signal,weak,vacuum}\right\} $
at the detector measuring the $i$ polarization state, where $i\in\left\{ H,V,D,A\right\} $ is: 
\begin{equation}
P_{\text{click}}^{i}\left(\mu_{k}\right)=1-(1-p_{bg}^{i})\cdot e^{-\eta_{SYS}^{i}\cdot\mu_{k}}\label{eq:PclickX}
\end{equation}
 The error probability is:
\begin{equation}
E^{i}\left(\mu_{k}\right)=1-(1-p_{bg}^{\perp i})\cdot e^{-e_{mis}\eta_{SYS}^{\perp i}\cdot\mu_{k}}\label{eq:ErrorX}
\end{equation}
where $\eta_{SYS}^{i}=\eta\times \eta_{BOB} \times \eta_{d}$ is the total transmission leading to detector $i$ (i.e. the channel transmittance $(\eta)$, the transmittance of Bob's optical instruments $(\eta_{BOB})$ and the
detector's quantum efficiency $(\eta_{d})$). In Eq.\ \eqref{eq:ErrorX}, $p_{bg}^{\perp i}$
is the background noise probability on the detector orthogonal to $i$ and $e_{mis}$ is the optical misalignment. We note
that the background noise probability is taken as a linear function of the channel's
transmittance $\eta$ : $p_{bg}=p_{bg}\left(\eta\right)=Y_{0}+b\cdot\eta$.

The numerical optimization returns the parameters $\left\{ q_{X},P_{\mu_1},P_{\mu_2},\mu_1,\mu_2\right\} $ that maximize the secure key rate $R=\dfrac{\ell}{N}$ for a given number $N$ of sent pulses ($N=3\times10^{10}$ for our experiment), where $\ell$
is the number of distilled bits \cite{Lim2014}, 
\begin{eqnarray}
\ell & = & \left\lfloor \vphantom{\frac{2}{\varepsilon_{cor}}} s_{X,0}+s_{X,1}\left(1-h(\phi_{X})\right)\right.\nonumber\\
 &  & \left.-n_{X} \! \cdot \! f_{EC} \! \cdot \!  h(e_{obs})-6\log_{2}\frac{21}{\varepsilon_{sec}}-\log_{2}\frac{2}{\varepsilon_{cor}}\right\rfloor \ \ \ \
\end{eqnarray}
To summarize the approach of Lim, \textit{et al.}, in \cite{Lim2014}, we estimate the lower bound of the zero-photon pulses contribution as:
\begin{equation}
    s_{X,0} \geq \tau_{0}\frac{\mu_2n_{X,\mu_3}^--\mu_3n_{X,\mu_2}^+}{\mu_2 - \mu_3}
\end{equation}
and the the lower bound of the single-photon pulses contribution as:
\begin{equation}
    s_{X,1} \geq \frac{\tau_1\mu_1[n_{X,\mu_2}^--n_{X,\mu_3}^+-\frac{\mu_2^2-\mu_3^2}{\mu_3^2}(n_{X,\mu_1}^+-\frac{s_{X,0}}{\tau_0})]}{\mu_1(\mu_2 - \mu_3) - \mu_2^2 + \mu_3^2}
\end{equation}

In the above we use the conditional probability $\tau_n$ that an $n-$photon pulse is sent:
\begin{equation}
    \tau_n = \sum_k \frac{\mathrm{e}^{-k}k^np_k}{n\mathrm{!}}
\end{equation}
and $n_{X,k}^{\pm}$ the number of detections where both Alice and Bob use the X basis, considering the finite sample size.
\begin{equation}
    n_{X,k}^{\pm} := \frac{\mathrm{e}^k}{p_{\mu_k}} \left[ n_{X,k} \pm \sqrt{\frac{n_X}{2}\mathrm{ln}\frac{21}{\varepsilon_{sec}}} \right]
\end{equation}
The detection numbers $n_{X,k}$ are calculated from Eqs.\ \eqref{eq:PclickX} and \eqref{eq:ErrorX}. Here $n_X=n_{X,\mu_1}+n_{X,\mu_2}+n_{X,\mu_3}$. The observed error in the rectilinear basis $e_{obs}$ is calculated as $e_{obs}=\frac{m_X}{n_X}$ with $m_X=m_{X,\mu_1}+m_{X,\mu_2}+m_{X,\mu_3}$. The numbers of errors $m_{X,k}$ are calculated from Eq.\ \eqref{eq:ErrorX}. Similar expressions hold in the diagonal basis by replacing $X\rightarrow Z$.

We estimate the upper bound of the phase error rate as:
\begin{equation}
    \phi_X \leq \frac{v_{Z,1}}{s_{Z,1}} + \gamma(\epsilon_{sec},\frac{v_{Z,1}}{s_{Z,1}},s_{Z,1},s_{X,1})
\end{equation}
Here $\gamma(\cdot)$ is the estimation uncertainty and $v_{Z,1}$ is the number of errors stemming from single photon pulses in the diagonal basis and is estimated as: 
\begin{equation}
    v_{Z,1} \leq \tau_{1}\frac{\mu_2m_{Z,\mu_2}^+-\mu_3m_{Z,\mu_3}^-}{\mu_2 - \mu_3}
\end{equation}
With $m_{Z,k}^{\pm}$ the number of errors in the diagonal basis considering the finite sample size.
\begin{equation}
    m_{Z,k}^{\pm} := \frac{\mathrm{e}^k}{p_{\mu_k}} \left[ m_{Z,k} \pm \sqrt{\frac{m_Z}{2}\mathrm{ln}\frac{21}{\varepsilon_{sec}}} \right]
\end{equation}


\section{Estimating the channel's transmittance with classical probe pulses}

\begin{figure}[h]
\begin{center}\includegraphics[scale=0.95]{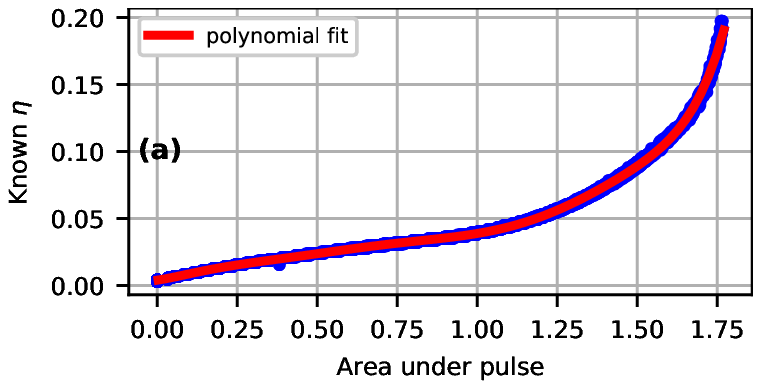}\end{center}
\begin{center}\includegraphics[scale=0.95]{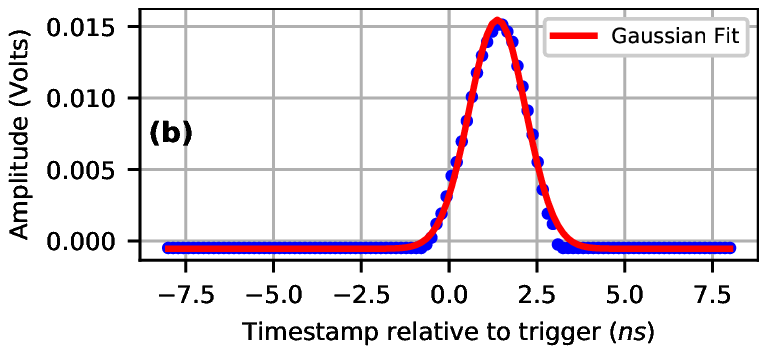}\end{center}
\caption{(a) Polynomial Fit to determine the correlation between the measured
area under the probe pulse and the programmed transmittance. (b) Example
of a probed pulse captured by the oscilloscope and its Gaussian fit.\label{fig:Probes}}
\end{figure}

In our experiment, classical probe pulses at a 4-kHz repetition rate, and $\sim$3 ns FWHM
at the 29 ITU channel are sent along the quantum pulses. After passing
the AOM, they are separated from the quantum pulses
with a DWDM and collected by a high-gain classical photo-detector.
We utilize the Fast-Frame feature of a DPO 7205 Tektronix Oscilloscope,
which stores samples in a short interval around the trigger
(16 ns in Figure \ref{fig:Probes}(b) sampled at 5 G-samples/sec).
Thus, we acquire high-resolution pulses (Figure \ref{fig:Probes}(b))
with minimum data storage. By performing a Gaussian fit on the pulses,
we acquire the area under each pulse, which is a direct measure of
the transmitted intensity. For an initial calibration set, we correlate
with a polynomial fit the measured pulse area with the programmed
transmittance. For the actual measurements, we use this polynomial
fit to deduce the transmittance given the measured pulse area. We note that we achieve similar resolution in Figure \ref{fig:Probes}(a) by simply calculating the sum of the samples of each frame, which is also significantly faster to compute compared to the Gaussian fits.

\end{document}